# Graphical abstract:

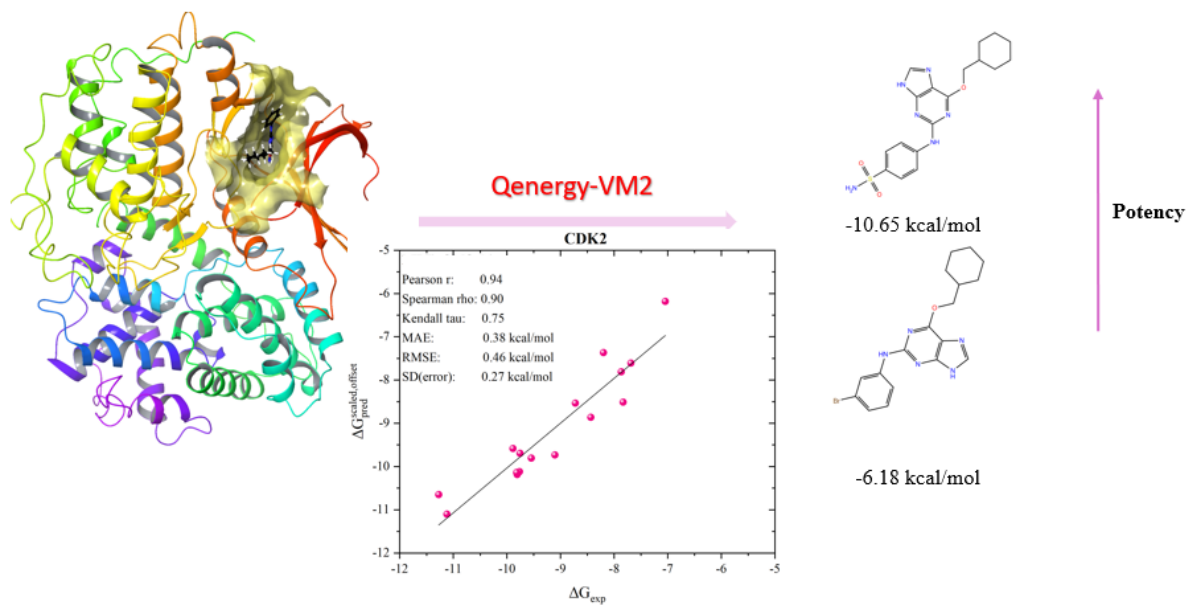

# Synergistic Computational Approaches for Accelerated Drug Discovery: Integrating Quantum Mechanics, Statistical Thermodynamics, and Quantum Computing

Farzad Molani[1] and Art E. Cho[1,2*]

[1] inCerebro, Co., Ltd., Seongdong-gu, Seoul, 04778, Republic of Korea

[2] Department of Bioinformatics, Korea University, 2511 Sejong-ro, Sejong, 30119, Republic of Korea

Email: artcho@korea.ac.kr

## Abstract

Accurately predicting protein–ligand binding free energies (BFEs) remains a central challenge in drug discovery, particularly because the most reliable methods, such as free energy perturbation (FEP), are computationally intensive and difficult to scale. Here, we introduce a hybrid quantum–classical framework that combines Mining Minima sampling with quantum mechanically refined ligand partial charges, QM/MM interaction evaluation, and variational quantum eigensolver (VQE)-based electronic energy correction. This design enables explicit treatment of polarization, charge redistribution, and electronic correlation effects that are often underestimated in purely classical scoring schemes, while retaining computational efficiency. Across 23 protein targets and 543 ligands, the method achieves a mean absolute error of ~1.10 kcal.mol$^{-1}$ with strong rank-order fidelity (Pearson R = 0.75, Spearman ρ = 0.76, Kendall τ = 0.57), consistent with the performance of contemporary FEP protocols. Notably, the workflow requires only ~25 minutes per ligand on standard compute resources, resulting in an approximate 20-fold reduction in computational cost relative to alchemical free energy approaches. This level of accuracy and efficiency makes the method well-suited for high-throughput lead optimization and iterative design cycles in pharmaceutical discovery. The framework also provides a natural foundation for future integration with machine learning models to enable predictive, large-scale, and adaptive screening strategies.

## 1. Introduction

The pharmaceutical industry continues to face escalating R&D costs and lengthy development timelines, highlighting a critical need for transformative advances in early-stage lead discovery and optimization [1]. Central to small-molecule drug discovery is the identification of ligands that bind strongly and selectively to therapeutic targets while maintaining favorable pharmacological properties. Accurate prediction of protein–ligand binding free energies (BFEs) plays a pivotal role in this process and has long been regarded as a key objective in computer-aided drug design (CADD). Among available computational strategies, physics-based free energy calculations based on Monte Carlo or molecular dynamics sampling provide the most rigorous theoretical foundation. However, classical approaches often encounter trade-offs between computational cost, accuracy, and scalability when confronted with the structural complexity and conformational diversity of biomolecular systems [2, 3]. Over the last decade, advances in sampling algorithms, force field accuracy, and computational efficiency have enabled routine application of relative free energy calculations in medicinal chemistry workflows [4-19]. Widely used simulation packages such as Schrodinger [7, 8], GROMACS [9], AMBER [20], and CHARMM [21], offer extensive support for Alchemical state transformation, λ-scheduling, and trajectory analysis, while preparation platforms such as CHARMM-GUI [22], Lomap [23], ProtoCaller [24], and FESetup [25] streamline system construction. Benchmarking studies demonstrate that FEP+ achieves mean absolute errors (MAEs) of 0.8–1.2 kcal·mol$^{-1}$ with correlations of R = 0.5–0.9 [7, 8], PMX yields R ≈ 0.3–1.0 depending on system characteristics [17], and Flare, MM-PBSA, and MM-GBSA typically perform with lower or more variable reliability [18, 19, 26]. These results underscore an ongoing challenge: achieving FEP-level predictive accuracy while reducing simulation expense to enable high-throughput application.

To address this gap, Gilson et al. benchmarked VeraChem's second-generation Mining Minima (MM-VM2) method, which estimates chemical potentials by systematically locating statistically relevant conformational minima [27]. MM-VM2 showed markedly higher accuracy compared to docking, MM-PBSA, and MM-GBSA, while exhibiting slightly lower performance than FEP+ (with a linear correlation lower by 0.2), yet it achieved this across 20 targets at a significantly reduced computational cost. Subsequent advancements incorporated quantum mechanical (QM) energetic corrections. Xu et al. introduced QM-VM2 for host–guest complexes, demonstrating improved agreement with experimental trends upon replacing MM energies with semi-empirical QM evaluations [28]. Building upon this direction and motivated by the success of our earlier efforts in incorporating quantum effects into protein–ligand modeling, we previously applied a QM/MM-based refinement protocol within docking pipelines [29-35]. This led to development of Qcharge-VM2, which integrates QM/MM-derived partial atomic charges into MM-VM2 sampling [36], followed by Qcharge-MC-FEPr, which performs free energy averaging across multiple QM/MM-refined conformers, achieving R ≈ 0.81 across 203 ligands [37]. More recently, Schlinsog et al. introduced PLQM-VM2 using DFTB3-D3(BJ)H/PCM corrections at ~35 CPU hours per complex [38], further demonstrating the potential of QM-refined MM-VM2 workflows.

Meanwhile, quantum computing is emerging as a complementary direction in CADD, with growing evidence of value in docking, affinity scoring, conformational exploration, and generative molecular design [39-41]. Among quantum-classical hybrid algorithms, the Variational Quantum Eigensolver (VQE) has shown particular promise for evaluating ground-state molecular energies in the NISQ era [42-44]. Recent studies have demonstrated that VQE-based interaction energy evaluation can reproduce experimental affinity trends in pharmaceutically relevant systems such as KDM5A, BACE1 and SARS-CoV-2 Mpro inhibitors [45-47].

Building on these developments, we introduce Qenergy-VM2, a refined free-energy workflow that integrates ESP-derived ligand charges, VM2-based conformational sampling, and a quantum-derived interaction energy correction step utilizing QM/MM and VQE-based energetic evaluation to better account for polarization, charge transfer, and electron correlation effects. The method applies these corrections to the statistically dominant binding conformers identified by the VM2 ensemble, yielding improved physical realism while maintaining computational efficiency.

The remainder of this manuscript is structured as follows: we first summarize the theoretical foundations of the MM-VM2 framework, then describe the Qenergy-VM2 workflow and the derivation of charge and interaction energy corrections. We then present benchmark results across 23 protein targets and 543 ligands, comparing Qenergy-VM2 to MM-GBSA, FEP+, PMX, and other methods. Finally, we discuss the implications for high-throughput drug discovery and outline future directions for integrating Qenergy-VM2 with machine learning–guided screening pipelines.

## 2. Methods

This section details the theoretical foundations, algorithmic structures, and practical implementations of the advanced computational frameworks that are reshaping molecular modeling and drug discovery. Each method is presented with a focus on its unique contributions, how it addresses existing challenges, and its potential for synergistic integration.

### 2.1. Quantum Mechanical Enhancements

This protocol (QMFF-VM2) enhances MM-VM2 by specifically improving how ligand atomic charges are treated. Rather than depending on the default force-field-assigned partial charges, we re-evaluate the ligand's charges using ESP values derived from QM calculations on the mapped/snapped conformers generated in the VM2 workflow. This substitution yields a more realistic description of the ligand's electronic environment and thus a more accurate representation of its interactions within the binding pocket.

### 2.2. The VM2 Based Free Energy Calculations

The theoretical background and methodological formulation of the VM2 framework have been extensively described in the literature and will not be repeated here [27, 36-38, 48]. In the present work, only the specific computational settings and modifications relevant to this study are outlined. The VM2 method calculates true free energies of binding by employing the "mining minima" approach, which is rigorously grounded in statistical thermodynamics [27, 48]. This approach is designed to circumvent the analytical intractability of the multi-dimensional configurational integral of the partition function. Instead of attempting to exhaustively sample the entire conformational space, which plagues traditional MD-based FEP methods due to their high computational cost, VM2 approximates this integral as a sum over local configuration integrals associated with a manageable set of low-energy minima of the system. This approach concentrates computational resources on the thermodynamically dominant low-energy states, ensuring a better trade-off between accuracy and efficiency. The standard BFE ($\Delta G^\circ_{bind}$) is determined from the standard chemical potentials (μ°) of the protein-ligand complex (PL), the free protein (P), and the free ligand (L), following the relationship: $\Delta G^\circ_{bind} = \mu^\circ_{PL} - \mu^\circ_P - \mu^\circ_L$. The chemical potential of each molecular species is computed by identifying its most stable local energy minima and extracting their contributions to the overall chemical potential. This process is iterative, with successive searches for low-energy minima and continuous calculation of a cumulative free energy. The calculation is considered converged when no new lower energy minima can be found, and the cumulative free energy stops changing significantly, ensuring that the most relevant conformational states have been identified and characterized. VM2 employs two primary search algorithms designed to efficiently explore the complex energy landscapes of biomolecular systems. First is rigid body translation/rotation search. This method systematically or randomly samples translations and rotations of the ligand relative to the protein. After each small step in translation or rotation, a few energy minimization steps are performed with respect to the non-ligand atoms in the system, allowing for relaxation in response to the ligand's distortion. Second is mode distort-minimize search. This technique involves random-pair or ligand-focused mode distortions. Cycles of distort-minimize operations use the previous lowest energy minimum as a basis for subsequent distortions, facilitating efficient exploration of conformational space. A key factor underlying VM2's accuracy is its explicit treatment of entropy in free energy calculations. VM2 accounts for entropic contributions using the harmonic approximation with mode scanning approach, while bulk solvation effects are incorporated through continuum models. When necessary, explicit water molecules can also be included in the binding site to more accurately represent localized solvation effects.

### 2.3. Hybrid Quantum–Classical Framework for Molecular Energy Calculations

Integrating QM/MM calculations with VQE forms a central component of our strategy to develop accurate and efficient computational workflows for CADD. By combining the atomistic detail of QM/MM with the quantum-level accuracy of VQE, our framework can simultaneously capture localized electronic interactions in the active site and long-range environmental effects from the protein matrix. Traditionally, QM/MM methodology has provided a practical balance for modeling large biomolecular systems: the chemically reactive region is treated quantum mechanically while the surrounding environment is described using molecular mechanics [49-51]. This partitioning substantially reduces computational cost while preserving chemical accuracy where it is most critical. In our workflow, the most probable ligand–protein binding pose generated by QMFF-VM2 conformational sampling is used as the structural input. The electronic structure of the free ligand is computed using VQE, while the ligand–protein complex is modeled within a QM/MM framework. This hybrid approach effectively leverages near-term quantum algorithms while mitigating current hardware constraints, enabling the study of large and pharmacologically relevant biomolecular systems within practical computational limits.

## 3. Computational Details

Figure 1 provides a schematic overview of the Qenergy-VM2 workflow. To improve electrostatic accuracy beyond fixed-charge force fields, ESP charges are derived from QM calculations on the free ligand. These calculations use conformers generated via the VM2 mapping/snapping procedure, performed with the PySCF engine [52] at the HF/madef2-svp [53] level, employing the ddCOSMO implicit solvent model for water. The Karlsruhe basis sets provide balanced and economical coverage from double- to quadruple-zeta quality for all elements up to radon, with minimally augmented (ma) variants including diffuse functions that, while less critical than in Pople sets, remain important for electron affinity and distribution [53]. The charge fitting process begins by evaluating the QM ESP at each point in a predefined grid located in the solvent-accessible region around the molecule [54]. These grid points are positioned outside the van der Waals radius of the molecule. The RESP fitting parameters include a probe radius of 0.7 Å, application of restraints to yield RESP atomic charges while excluding hydrogen atoms from the restraint procedure, with restraint parameters alpha=0.001 au and beta=0.1 au, a maximum of 25 fitting iterations, and a convergence tolerance of $1.0 \times 10^{-4}$ electrons.

After rescaling the ESP partial charges, a rigorous conformational search is performed. Solvation effects during the minima search are modeled with the Generalized Born implicit solvent, and final BFE evaluations are carried out using Poisson–Boltzmann (PB) Surface Area calculations [27]. The dielectric constants are set to 1 for the protein interior and 80 for the solvent exterior, with the Richards molecular surface used as the dielectric boundary. The ligand and binding-site residues within 4 Å are treated as flexible ("live" set), while atoms within 6 Å beyond this live region are fixed but included in energy calculations ("real" set) [27].

Non-covalent interactions, such as hydrogen bonding, halogen bonding, ionic, fluorine bonding, cation–π, and π–π interactions, are explicitly evaluated, with the enthalpic term expressed as ΔH = ΔU + ΔW. Here, ΔU includes valence, Coulombic, and Lennard-Jones contributions, while ΔW accounts for polar and nonpolar solvation components. The BFEs will be calculated using the following equation:

$$\Delta G = \Delta U_{MM} + \Delta W - T\Delta S \qquad (1)$$

Where

$$\Delta W = \Delta E_{PB} + \Delta E_{SA} \qquad (2)$$
$$\Delta U_{MM} = \Delta E_{val} + \Delta E_{coul} + \Delta E_{vdw} \qquad (3)$$

W and $U_{MM}$, $\Delta E_{PB}$, $\Delta E_{SA}$, $\Delta E_{val}$, $\Delta E_{coul}$, $\Delta E_{vdw}$, -TΔS, ΔG are solvation energy, classical internal potential energy, change in mean Poisson-Boltzmann solvation energy, change in nonpolar solvation energy, change in mean energy associated with force-field bond stretch, angle bend, and dihedral terms, change in mean force field Coulombic energy, change in mean force field Lenard-Jones energy, change in configurational entropy contribution to the free energy, and BFE, respectively. The VQE+QM/MM protocol was developed to enhance the internal energy term within the QMFF-VM2 free energy calculation. QM/MM and VQE are implemented using PySCF [52] and Tangelo platform [55], respectively. Quantum calculations are performed on a 10-qubit emulator (6 active electrons) employing a MINAO basis set and the unitary coupled-cluster singles and doubles (UCCSD) ansatz. The classical optimizer used is SLSQP [56], with symmetry-conserving Bravyi-Kitaev (scBK) mapping [57] employed for qubit mapping. In the QM/MM framework, the ligand was treated quantum mechanically while the surrounding protein environment was modeled using a classical force field. This partitioning allows the method to capture the essential electronic correlation and polarization effects that govern binding interactions, while maintaining computational efficiency for large biomolecular systems. Partial atomic charges for the protein were generated using tleap with the ff99SB force field [58], and these charges were used to construct the MM region. The QM region was evaluated using the B3LYP-D4 exchange–correlation functional to include empirical dispersion corrections [59]. A def2-SVPD basis set was applied to all QM atoms, and the def2-universal-jfit auxiliary basis was employed to enable density-fitting (RI-DFT) acceleration [60]. To avoid artificial over-polarization effects and reduce computational cost, only protein residues with at least one atom located within 8.0 Å of the ligand were included in the MM region. These MM atoms were treated as fixed point charges, extracted directly from the topology and coordinate files. Their electrostatic influence on the QM region was incorporated using the electrostatic embedding approach, where the MM charge distribution is included explicitly in the one-electron part of the QM Hamiltonian [61]. The (electronic) binding energy, $U_{bind}$, is chosen as a simple metric by which to measure the strength of the interaction between the protein and the ligand, and is calculated as

$$\Delta U_{bind} = U_{complex} - E_{ligand} - E_{protein} \qquad (4)$$

Mishra et al. introduced an additional QM term to refine the BFE by incorporating QM/MM and MM-GBSA contributions into the total BFE [62]. Building on this idea and employing a "frozen protein" approximation [46, 47], the protein is treated as a set of distributed point charges, allowing the protein–protein terms in $U_{complex}$ to cancel and simplifying the expression to a single energy difference.

$$\Delta U_{QM/MM}^{VQE} = U_{ligand\_in\_protein} - U_{ligand} \qquad (5)$$

$$\Delta G = \Delta U_{MM} + \Delta W - T\Delta S + \alpha \Delta U_{QM/MM}^{VQE} \qquad (6)$$

Where α is a constant value of 1.59 ×10$^{-3}$.

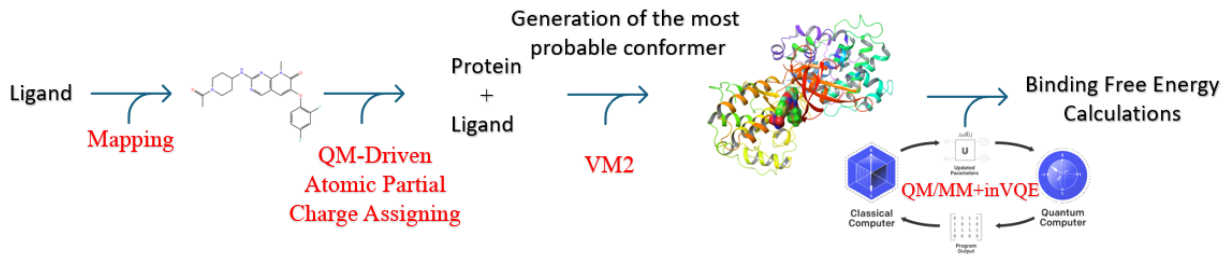

**Figure 1** Generalized framework for Qenergy-VM2.

4. **Computational cost**

The computational cost of FEP methods remains substantial even when using parallel clusters. For instance, in Flare, BFE calculations for a ligand pair in TYK2 with 9 λ windows over a 4 ns simulation required roughly 15 hours on a single GPU [19]. Li et al. reported that absolute BFE calculations took approximately 10 hours per ligand on eight NVIDIA GeForce GTX-580 GPUs [26]. Frush et al. combined MD simulations and QM/MM calculations on a single CPU, requiring about 20 hours for a single conformer [51]. The Schrödinger team reported that FEP+ required approximately 7.38 days on an NVIDIA Tesla V100 GPU to run 58 perturbations for the BACE target (36 ligands) [7, 62]. By comparison, our Qenergy-VM2 protocol completes calculations in about 15 h on an Intel(R) Xeon(R) CPU E5-2680 v3 at 2.50 GHz, using 64 MPI processes.

## 5. Results and Discussion

### 5.1. Target-Wise Evaluation of BFE Prediction Performance

This section presents the quantitative performance of a synergistic computational approach for predicting BFEs across a diverse set of 23 protein targets, encompassing over 543 unique ligands. The overall agreement between experimental and predicted values will be evaluated using Pearson, Spearman, and Kendall correlation coefficients, along with MAE, root mean square error (RMSE), and standard deviation (SD) for the datasets with a direct comparison against a state-of-the-art FEP (FEP+ [8, 15, 16], PMX [17], AMBER20 [18], Flare [19]) benchmark where comparative data is available. The quality of the calculated BFEs can be evaluated through the correlation between the predicted $\Delta G_{calc}^{offset,scaled}$, as defined earlier [37]. We found that a "universal scaling factor" (USF) of 0.50 minimizes the error of the predicted values relative to the experimental ones, without introducing an unreasonable slope in the correlation.

For systematic comparison, the targets were grouped into four mechanistic classes based on binding-site architecture and functional characteristics. Group I (Proteases) includes HIV-1 protease (38 neutral ligands), BACE (36 charged ligands), BACE-Hunt (32 charged ligands), BACE-P2 (16 neutral ligands), and Thrombin (10 charged ligands). These proteins contain deep, well-defined catalytic pockets and are widely used as benchmarks for assessing scoring accuracy. Group II (Protein Kinases and Signaling Enzymes) comprises CDK2 (16 neutral ligands), JNK1 (21 neutral ligands), p38α (34 neutral ligands), TYK2 (16 neutral ligands), c-MET (24 mixed ligands), PTP1B (23 charged ligands), SYK (38 neutral ligands), and EG5 (28 mixed ligands). These targets have conformationally flexible ATP-binding clefts. Group III (Shallow or Induced-Fit Pockets) includes MCL1 (42 charged ligands), PDE2 (20 neutral ligands), TNKS2 (21 neutral ligands), PFKB3 (40 neutral ligands), and HIF-2α (42 neutral ligands), in which ligand binding is associated with surface-exposed or inducible pockets requiring local structural rearrangement. Group IV (G Protein–Coupled Receptors) consists of A2A (Minetti: 9 neutral ligands; Piersanti: 7 neutral ligands), β1-adrenergic receptor (9 neutral ligands), δ-opioid receptor (11 neutral ligands), and CXCR4 (10 charged ligands) representing membrane-embedded, highly dynamic systems that present distinct challenges for binding free-energy prediction. All structural and affinity data were obtained from the Open Force Field protein–ligand benchmark (https://github.com/openforcefield/proteinligand-benchmark), Gilson et al. [29] , and Lenselink et al. [16]. In the Supplementary Information (Figure S1), we provide the correlation plot of predicted versus experimental ΔG values for the full dataset evaluated with the Qenergy-VM2 protocol, along with target-specific absolute error (offset-scaled) distribution histograms.

Table 1 Statistical performance comparison of various computational methods in predicting Group I dataset binding free energies, showing correlation (R, ρ, τ), error (MAE, RMSE), and deviation (SD).

| HIV 1 Protease | R | ρ | τ | MAE | RMSE | SD | Slope |
|---|---|---|---|---|---|---|---|
| Qenergy-VM2 | 0.83 | 0.66 | 0.48 | 0.88 | 1.11 | 0.67 | 0.66 |
| **BACE** | R | ρ | τ | MAE | RMSE | SD | Slope |
| Qenergy-VM2 | 0.69 | 0.61 | 0.44 | 0.62 | 0.77 | 0.46 | 0.92 |
| FEP(R) | AMBER20 | FEP+ | PMX-GAFF | | PMX-CGenFF | | Flare |
| | 0.54 | 0.66 | 0.45 | | 0.46 | | 0.44 |
| **BACE (hunt)** | R | ρ | τ | MAE | RMSE | SD | Slope |
| Qenergy-VM2 | 0.59 | 0.52 | 0.38 | 1.13 | 1.47 | 0.94 | 0.92 |
| FEP(R) | AMBER20 | FEP+ | PMX-GAFF | | PMX-CGenFF | | Flare |
| | --- | 0.85 | 0.75 | | 0.69 | | --- |
| **BACE (p2)** | R | ρ | τ | MAE | RMSE | SD | Slope |
| Qenergy-VM2 | 0.91 | 0.74 | 0.54 | 0.64 | 0.75 | 0.40 | 1.47 |
| FEP(R) | AMBER20 | FEP+ | PMX-GAFF | | PMX-CGenFF | | Flare |
| | --- | 0.60 | 0.67 | | 0.37 | | --- |
| **Thrombin** | R | ρ | τ | MAE | RMSE | SD | Slope |
| Qenergy-VM2 | 0.60 | 0.76 | 0.60 | 1.23 | 1.81 | 1.32 | 2.29 |
| FEP(R) | AMBER20 | FEP+ | PMX-GAFF | | PMX-CGenFF | | Flare |
| | 0.91 | 0.45 | 0.11 | | -0.09 | | 0.31 |

Statistical performance for Group I targets is summarized in Table 1, reporting correlation metrics (R, ρ, τ), error measures (MAE, RMSE), SD, and regression slopes. For HIV-1 protease, Qenergy-VM2 exhibited strong correlation and high precision. The linear correlation coefficient was 0.83, with rank correlations of ρ = 0.66 and τ = 0.48, indicating good agreement with the experimental rank ordering. Quantitative accuracy was also notable, with an MAE of 0.88 kcal·mol⁻¹, an RMSE of 1.11 kcal·mol⁻¹, and a standard deviation of 0.67 kcal·mol⁻¹. The regression slope of 0.66 suggests a consistent but slightly compressed energy scale. On the BACE dataset, Qenergy-VM2 delivered stronger predictive performance than other approaches, yielding higher correlations alongside a solid MAE of 0.62 kcal.mol⁻¹ and an RMSE of 0.77 kcal.mol⁻¹. It also maintained a low SD (0.46) and regression slopes close to unity. Compared with reported results for AMBER20 (R = 0.54), FEP+ (R = 0.66), PMX (R = 0.45–0.46), and Flare (R = 0.44), Qenergy-VM2 offers a compelling combination of accuracy and robustness. For the more challenging BACE (Hunt) subset, characterized by highly flexible ligands, Qenergy-VM2 achieved reasonable rank correlations (R = 0.59, ρ = 0.52, τ = 0.38) and reduced errors (MAE = 1.13 and RMSE = 1.47). The regression slope was close to unity, highlighting accurate energetic scaling. While FEP methods can reach correlations of R = 0.69–0.85, Qenergy-VM2 attains comparable accuracy at a fraction of the computational cost, demonstrating its efficiency for systems involving substantial ligand flexibility. For the BACE (p2) subset, Qenergy-VM2 achieved excellent correlation (R = 0.91) and strong rank-order metrics (ρ = 0.74, τ = 0.54), outperforming reported FEP variants (R = 0.37–0.67). With MAE and RMSE both below 0.75 kcal.mol$^{-1}$, Qenergy-VM2 demonstrates high predictive precision. These results indicate that Qenergy-VM2 delivers stable ligand ranking while maintaining strong numerical accuracy. For the Thrombin dataset, Qenergy-VM2 achieved R = 0.60, ρ = 0.76, and τ = 0.60, with an MAE of 1.23 kcal.mol$^{-1}$ and a regression slope of 2.29. Among alchemical techniques, AMBER20 performed best (R = 0.91), whereas FEP+ showed weaker correlation (R = 0.45), PMX-based methods performed poorly (R ≈ –0.1 to 0.1), and Flare also showed limited accuracy (R = 0.31). Overall, Qenergy-VM2 provides stable, mid-range accuracy, surpassing most reported FEP outcomes, while preserving coherent ranking behavior and uniform error distribution across the ligand set.

**Table 2** Statistical performance comparison of various computational methods in predicting Group II benchmark dataset binding free energies, showing correlation (R, ρ, τ), error (MAE, RMSE), and deviation (SD).

| CDK2 | R | ρ | τ | MAE | RMSE | SD | Slope |
|---|---|---|---|---|---|---|---|
| Qenergy-VM2 | 0.94 | 0.90 | 0.75 | 0.38 | 0.46 | 0.27 | 1.04 |
| FEP(R) | AMBER20 | FEP+ | | PMX-GAFF | PMX-CGenFF | | Flare |
| | 0.45 | 0.71 | | 0.65 | 0.48 | | 0.73 |
| JNK1 | R | ρ | τ | MAE | RMSE | SD | Slope |
| Qenergy-VM2 | 0.77 | 0.76 | 0.63 | 1.33 | 1.46 | 0.62 | 1.82 |
| FEP(R) | AMBER20 | FEP+ | | PMX-GAFF | PMX-CGenFF | | Flare |
| | 0.56 | 0.79 | | 0.51 | 0.52 | | 0.72 |
| P38α | R | ρ | τ | MAE | RMSE | SD | Slope |
| Qenergy-VM2 | 0.70 | 0.71 | 0.54 | 0.77 | 0.95 | 0.56 | 0.95 |
| FEP(R) | AMBER20 | FEP+ | | PMX-GAFF | PMX-CGenFF | | Flare |
| | 0.75 | 0.76 | | 0.72 | 0.68 | | 0.63 |
| TYK2 | R | ρ | τ | MAE | RMSE | SD | Slope |
| Qenergy-VM2 | 0.82 | 0.70 | 0.57 | 0.73 | 0.98 | 0.65 | 1.11 |
| FEP(R) | AMBER20 | FEP+ | | PMX-GAFF | PMX-CGenFF | | Flare |
| | 0.58 | 0.86 | | 0.60 | 0.58 | | 0.73 |
| c-Met | R | ρ | τ | MAE | RMSE | SD | Slope |
| Qenergy-VM2 | 0.87 | 0.85 | 0.69 | 0.98 | 1.16 | 0.61 | 1.16 |
| FEP(R) | AMBER20 | FEP+ | | PMX-GAFF | PMX-CGenFF | | Flare |
| | --- | 0.91 | | 0.78 | 0.84 | | --- |
| PTP1B | R | ρ | τ | MAE | RMSE | SD | Slope |
| Qenergy-VM2 | 0.79 | 0.62 | 0.50 | 1.74 | 2.18 | 1.31 | 1.86 |
| FEP(R) | AMBER20 | FEP+ | PMX-GAFF | | PMX-CGenFF | | Flare |
| | 0.72 | 0.74 | 0.71 | | 0.54 | | 0.69 |
| SYK | R | ρ | τ | MAE | RMSE | SD | Slope |
| Qenergy-VM2 | 0.27 | 0.27 | 0.17 | 0.93 | 1.24 | 0.83 | 0.43 |
| FEP(R) | FEP+ | | | | | | |
| | 0.50 | | | | | | |
| EG5 | R | ρ | τ | MAE | RMSE | SD | Slope |
| Qenergy-VM2 | 0.50 | 0.38 | 0.23 | 1.65 | 2.09 | 1.28 | 1.34 |
| FEP(R) | FEP+ | | | | | | |
| | 0.71 | | | | | | |

Table 2 summarizes the statistical performance for the Group II targets. For CDK2, Qenergy-VM2 delivered exceptionally high predictive accuracy, achieving R = 0.94, ρ = 0.90, τ = 0.75, and excellent amount of error (MAE = 0.38 and RMSE = 0.46 kcal.mol$^{-1}$). It also maintained a low SD (0.27 kcal.mol$^{-1}$) and produced a broadened yet well-calibrated regression slope of 1.04. Compared with alchemical approaches, Qenergy-VM2 consistently outperformed AMBER20 and PMX variants and matched or exceeded typical FEP+ accuracy, all while requiring substantially lower computational cost. For JNK1, Qenergy-VM2 showed strong correlation (R = 0.77, ρ = 0.76, τ = 0.63), though the absolute errors were reasonable (MAE = 1.33 kcal·mol$^{-1}$; slope = 1.82), suggesting accurate trend capture but an over-expanded energy scale. FEP+ and Flare achieved comparable correlations (R ≈ 0.72–0.79) but required substantially greater sampling effort. These results indicate that Qenergy-VM2 effectively reproduces binding trends, though further calibration may be beneficial for systems with pronounced ligand-induced polarization. For p38α, Qenergy-VM2 demonstrated acceptable ranking fidelity (R = 0.70, ρ = 0.71, τ = 0.54) and a physically realistic energy scaling (slope = 0.95), with errors in line with expectations (MAE ≈ 0.77 kcal·mol$^{-1}$). Its performance was comparable to reported FEP-based methods (R ≈ 0.63–0.76), while requiring substantially less computational effort. For TYK2, Qenergy-VM2 achieved strong correlations (R = 0.82, ρ = 0.70, τ = 0.57) with an MAE of 0.73 kcal.mol$^{-1}$ and a near-unity slope (1.11). Compared with FEP+, which performed slightly better (R = 0.86), Qenergy-VM2 offers competitive predictive accuracy while being significantly more computationally efficient. For c-MET, Qenergy-VM2 displayed also strong correlation (R = 0.87, ρ = 0.85, τ = 0.69) with reasonable error (MAE = 0.98 kcal·mol$^{-1}$) and slope (1.16), indicating enhanced ranking discrimination but slightly exaggerated energetic spread. Its performance remains competitive with FEP+ (R = 0.91) and surpasses PMX-based methods. For PTP1B, Qenergy-VM2 showed linear correlation (R = 0.79), though the MAE (1.74 kcal·mol$^{-1}$) and steep slope (1.86) suggest strong ranking with over-amplified energy differences. This reflects the highly solvent-exposed, charge-sensitive active site, where additional sampling may improve accuracy. Its performance is comparable to the upper range of reported FEP+ results. For SYK, Qenergy-VM2 produced poor correlation (R = 0.27, slope = 0.43) with MAE = 0.93 kcal·mol⁻¹ and RMSE = 1.24 kcal·mol⁻¹, indicating that electronic polarization plays a critical role in this kinase and is better captured by QM refinement. FEP+ performed better (R ≈ 0.50) but at a higher computational cost. For EG5, Qenergy-VM2 exhibited limited predictive accuracy (R = 0.50; MAE = 1.65 kcal·mol⁻¹; RMSE = 2.09 kcal·mol⁻¹; SD = 1.34 kcal·mol⁻¹), whereas FEP+ achieved stronger correlation (R = 0.71). This suggests that EG5 remains a challenging target, where conformational flexibility and parameterization limitations affect performance across scoring frameworks.

**Table 3** Statistical performance comparison of various computational methods in predicting Group III benchmark dataset binding free energies, showing correlation (R, ρ, τ), error (MAE, RMSE), and deviation (SD).

| MCL1 | R | ρ | τ | MAE | RMSE | SD | Slope |
|---|---|---|---|---|---|---|---|
| Qenergy-VM2 | 0.66 | 0.68 | 0.46 | 1.33 | 1.76 | 1.15 | 1.42 |
| FEP(R) | AMBER20 | FEP+ | PMX-GAFF | | PMX-CGenFF | | Flare |
| | 0.51 | 0.73 | 0.42 | | 0.15 | | 0.74 |
| PDE2 | R | ρ | τ | MAE | RMSE | SD | Slope |
| Qenergy-VM2 | 0.66 | 0.66 | 0.55 | 1.15 | 1.60 | 1.11 | 1.62 |
| FEP(R) | FEP+ | PMX-GAFF | | PMX-CGenFF | | | |
| | 0.54 | 0.54 | | 0.57 | | | |
| TNKS2 | R | ρ | τ | MAE | RMSE | SD | Slope |
| Qenergy-VM2 | 0.46 | 0.43 | 0.26 | 0.91 | 1.08 | 0.58 | 0.49 |
| FEP(R) | FEP+ | | | | | | |
| | 0.40 | | | | | | |
| PFKB3 | R | ρ | τ | MAE | RMSE | SD | Slope |
| Qenergy-VM2 | 0.58 | 0.65 | 0.44 | 1.73 | 2.40 | 1.66 | 1.53 |
| FEP(R) | FEP+ | | | | | | |
| | 0.79 | | | | | | |
| HIF2a | R | ρ | τ | MAE | RMSE | SD | Slope |
| Qenergy-VM2 | 0.51 | 0.47 | 0.33 | 1.38 | 1.81 | 1.16 | 0.95 |
| FEP(R) | FEP+ | | | | | | |
| | 0.61 | | | | | | |

Table 3 summarizes the statistical performance for the Group III targets. For the MCL1 ligand series, Qenergy-VM2 shows solid correlation and rank-ordering (R = 0.66, ρ = 0.68, τ = 0.46), with a reasonable MAE (1.33 kcal·mol⁻¹) but a somewhat elevated RMSE (1.76 kcal·mol⁻¹) and a modest slope (1.42). In comparison, alchemical FEP methods display wide variability: FEP+ and Flare achieve stronger agreement with experiment (R ≈ 0.73–0.74), whereas AMBER20 and PMX-based workflows perform less favorably (R = 0.15–0.51). These trends reflect the complex interplay of polarization and dispersion within the MCL1 binding pocket. For PDE2, Qenergy-VM2 delivers consistent correlation and rank-ordering (R = 0.66, ρ = 0.66, τ = 0.55) together with a reasonable MAE (1.15 kcal·mol$^{-1}$), despite a relatively high RMSE and slope (1.62). This indicates that Qenergy-VM2 effectively separates strong and weak binders, though with an expanded energetic range. Available FEP benchmarks (R ≈ 0.54–0.57) suggest that Qenergy-VM2 provides superior ligand ranking for this target. For the TNKS2 series, Qenergy-VM2 shows moderate performance (R = 0.46, ρ = 0.43, τ = 0.26), with a good MAE (0.91 kcal·mol$^{-1}$) but a compressed slope (0.49), reflecting partial recovery of energetic trends across ligands. Notably, FEP+ also performs only moderately (R ≈ 0.40), consistent with the narrow experimental affinity range and conformational sensitivity characteristic of TNKS2. For PFKB3, Qenergy-VM2 demonstrates good rank-ordering (R = 0.58, ρ = 0.65) with a slightly elevated MAE (1.73 kcal·mol$^{-1}$) and a high slope (1.53), suggesting amplified scaling of predicted affinity differences. FEP+ exhibits stronger performance (R ≈ 0.79), in line with prior studies highlighting the electrostatic heterogeneity of this binding site. For the HIF2α ligand set, Qenergy-VM2 achieves good correlation and ranking (R = 0.51, ρ = 0.47, τ = 0.33) with reasonable error (MAE = 1.38 kcal·mol$^{-1}$). Available FEP+ results (R = 0.61) indicate moderate expected performance on this challenging target. These findings suggest that Qenergy-VM2 is limited by the subtle energy landscape of HIF2α, where small conformational and solvation effects strongly modulate affinity.

**Table 4** Statistical performance comparison of various computational methods in predicting group IV benchmark data set binding free energies, showing correlation (R, ρ, τ), error (MAE, RMSE), deviation (SD), and regression slope.

| A2a (Minetti et al.) | R | ρ | τ | MAE | RMSE | SD | Slope |
|---|---|---|---|---|---|---|---|
| Qenergy-VM2 | 0.41 | 0.30 | 0.28 | 0.82 | 0.99 | 0.56 | 0.39 |
| MM-GBSA | -0.31 | | | | | | |
| FEP+ | 0.78 | | | | | | |
| A2a (Piersanti et al.) | R | ρ | τ | MAE | RMSE | SD | Slope |
| Qenergy-VM2 | 0.55 | 0.39 | 0.24 | 0.50 | 0.57 | 0.27 | 0.83 |
| MM-GBSA | 0.33 | | | | | | |
| FEP+ | 0.55 | | | | | | |
| CXCR4 | R | ρ | τ | MAE | RMSE | SD | Slope |
| Qenergy-VM2 | 0.89 | 0.84 | 0.67 | 0.48 | 0.71 | 0.52 | 0.98 |
| MM-GBSA | 0.11 | | | | | | |
| FEP+ | 0.45 | | | | | | |
| δ-Opioid | R | ρ | τ | MAE | RMSE | SD | Slope |
| Qenergy-VM2 | 0.84 | 0.85 | 0.67 | 1.75 | 2.14 | 1.22 | 2.61 |
| MM-GBSA | -0.25 | | | | | | |
| FEP+ | 0.85 | | | | | | |
| β1-adernergic | R | ρ | τ | MAE | RMSE | SD | Slope |
| Qenergy-VM2 | 0.80 | 0.78 | 0.56 | 0.85 | 1.17 | 0.80 | 1.87 |
| MM-GBSA | 0.64 | | | | | | |
| FEP+ | 0.39 | | | | | | |

All Group IV targets were taken from the Lenselink et al. study [16], and the corresponding results are summarized in Table 4. For the A2A Minetti dataset, Qenergy-VM2 produced moderately accurate rank-ordering (R = 0.41, ρ = 0.30, τ = 0.28) with favorable error metrics (MAE = 0.82 kcal·mol$^{-1}$; RMSE = 0.99 kcal·mol$^{-1}$). Previously published MM-GBSA results for this benchmark reported negative correlation (R ≈ –0.31). Although FEP+ reached substantially higher accuracy (R ≈ 0.78), this comes at much greater computational cost. The low Qenergy-VM2 slope (0.39) indicates recovery of a physically meaningful binding-energy scale, helping narrow the gap between classical scoring approaches and resource-intensive alchemical methods. For the A2A Piersanti dataset, Qenergy-VM2 achieved good correlation (R = 0.55, ρ = 0.39, τ = 0.24), low prediction error (MAE = 0.50 kcal·mol$^{-1}$; RMSE = 0.57 kcal·mol$^{-1}$), and accurate scaling (slope = 0.83). MM-GBSA again delivered weak rank-ordering (R = 0.33), while FEP+ reached comparable correlation (R = 0.55) but required dramatically higher computational resources. These findings highlight Qenergy-VM2 as an efficient alternative capable of FEP-like predictive performance at a fraction of the cost. For the CXCR4 ligand series, Qenergy-VM2 provided excellent rank-ordering (R = 0.89, ρ = 0.84, τ = 0.67) with low error (MAE = 0.48 kcal·mol$^{-1}$) and a nearly ideal slope (0.98), indicating well-calibrated energy scaling. MM-GBSA performed poorly (R = 0.11), and FEP+ showed only moderate accuracy (R = 0.45). These results demonstrate that Qenergy-VM2 delivers high accuracy while being far more computationally efficient than alchemical methods. For the δ-opioid dataset, Qenergy-VM2 produced strong rank correlation (R = 0.84, ρ = 0.85, τ = 0.67) but higher prediction errors (MAE = 1.75 kcal·mol$^{-1}$; RMSE = 2.14 kcal·mol$^{-1}$) and a steep slope (2.61). This suggests that although Qenergy-VM2 show hich correlation and ranking, δ-opioid binding likely requires additional solvation or entropic treatments to reduce over-scaling. MM-GBSA again performed poorly (R = –0.25), while FEP+ showed high correlation (R = 0.85) but with high computational expense. Overall, Qenergy-VM2 provides reliable ligand ranking supported by a more physically grounded energy model. For the $β_1$-adrenergic receptor, Qenergy-VM2 showed strong ranking accuracy (R = 0.80, ρ = 0.78, τ = 0.56) with good errors (MAE = 0.85 kcal·mol$^{-1}$; RMSE = 1.17 kcal·mol$^{-1}$) and moderated energy scaling (slope = 1.87). MM-GBSA performed less favorably (R = 0.64), while FEP+ yielded weak correlation (R = 0.39). These results indicate that Qenergy-VM2 effectively corrects systematic overestimation seen in MM-based scoring while maintaining computational efficiency.

## 5.2. Overall Performance Across All Targets

Across all receptor datasets, Qenergy-VM2 consistently delivers strong correlation and rank-ordering with reasonable prediction errors, and in many cases approaches or matches the accuracy of FEP+. Although FEP+ remains the gold standard for rigorous free-energy estimation, its performance can be highly sensitive to sampling convergence and conformational restraints. In contrast, Qenergy-VM2 achieves comparable correlation, MAE, and RMSE at a fraction of the computational cost, making it an appealing quantum-enhanced alternative for early-stage lead optimization. To quantify the impact of QM refinement, we evaluated Qenergy-VM2 across a benchmark of 543 ligands spanning 23 protein targets (Table 5). The mean R-value reaches 0.68, with mean $\rho$ and $\tau$ values of 0.63 and 0.48, respectively, demonstrating reliable ligand ranking. MAE and RMSE remain within a reasonable range, indicating that the primary improvements stem from enhanced ranking and energetic discrimination rather than uniform scaling corrections. Based on reported data for FEP+, the mean R-value across the same dataset is approximately 0.67 with an RMSE of ~1.1 kcal·mol$^{-1}$. These results show that our approach is comparable to FEP+ in both accuracy and precision, while requiring significantly fewer computational resources. Notably, the mean regression slope of 1.24 remains close to unity, reflecting suitable differentiation between strong and weak binders while maintaining physical proportionality to experimental free-energy scales.

Figure 2 shows the predicted versus experimental BFEs for Qenergy-VM2 across the full dataset. Analysis of the full dataset using the Qenergy-VM2 protocol demonstrates strong predictive performance across multiple statistical metrics. For the complete set of ligands, R, $\rho$, and $\tau$ correlations reach 0.75, 0.76, and 0.57, respectively, indicating robust linear and rank-order agreement with experimental binding free energies. The MAE and RMSE are 1.10 and 1.52 kcal·mol$^{-1}$, respectively, with a regression slope of 1.04, confirming that the predicted values closely track experimental trends. To assess the influence of extreme deviations, the top 5% of outliers were removed, resulting in 27 data points being excluded. Following this adjustment, all correlation measures improved, with R, $\rho$, and $\tau$ values rising to 0.81, 0.80, and 0.61, respectively. Concurrently, error metrics decreased substantially (MAE = 0.94 kcal·mol$^{-1}$; RMSE = 1.19 kcal·mol$^{-1}$), and the regression slope reached unity (1.00) with a near-zero intercept (−0.09), indicating excellent proportionality between predicted and experimental binding free energies.

Importantly, the Kendall τ ≥ 0.5 threshold proposed by Schindler et al. [15], a criterion for "good" rank-ordering performance in pharmaceutical benchmarks, is satisfied, supporting the suitability of Qenergy-VM2 for ligand prioritization during lead optimization. Supplementary Figure S2 provides additional residual analyses for the full dataset, where residual plots including residuals versus predictors, residual histograms, residuals versus fitted values, and normal probability plots confirm normality, constant variance, and independence of errors. These results highlight that Qenergy-VM2 not only provides accurate and consistent predictions but also exhibits resilience to extreme values and satisfies linear-model assumptions, supporting its application in lead optimization and high-throughput screening workflows.

**Table 5** Mean Statistical performance of Qenergy-VM2 and FEP+ across all studied systems.

| Mean | R | ρ | τ | MAE | RMSE | SD | Slope |
|---|---|---|---|---|---|---|---|
| **Qenergy-VM2** | 0.68 | 0.63 | 0.48 | 1.04 | 1.33 | 0.82 | 1.24 |
| **FEP+** | 0.67 | --- | --- | --- | --- | --- | --- |

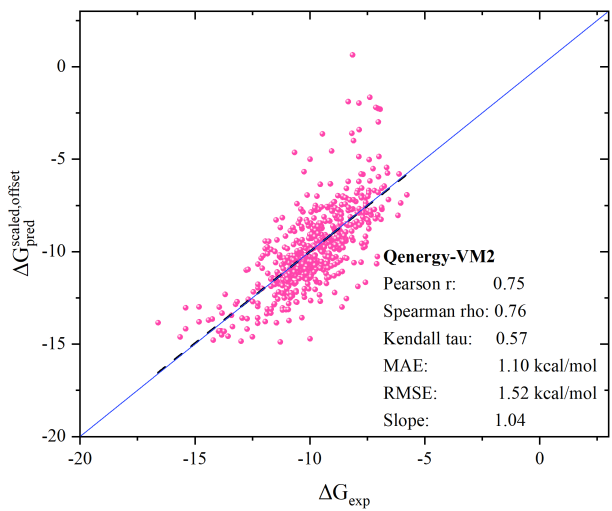

**Figure 2** Predicted versus experimental BFEs for the full ligand set using Qenergy-VM2. Corresponding statistical performance metrics are reported within plot.

## 6. Conclusions

In this work, we introduced the Qenergy-VM2 framework, which combines QM, statistical thermodynamics, and quantum computing to account for electronic polarization and charge-redistribution effects through QM-derived energetic corrections. Evaluated across 543 ligands spanning 23 protein targets, Qenergy-VM2 demonstrated strong predictive performance, achieving

R, ρ, and τ correlations of 0.75, 0.76, and 0.57 for the full dataset, with MAE and RMSE of 1.10 and 1.52 kcal·mol⁻¹, respectively. These results confirm the suitability of Qenergy-VM2 for reliable ligand prioritization during lead optimization. Statistical analyses further validated the predicted energies, confirming the assumptions of normality, constant variance, and independence of errors, which underscores the robustness and consistency of the protocol. A key practical advantage of Qenergy-VM2 is its computational efficiency: the method requires approximately 25 minutes per ligand, making it roughly 20 times faster than conventional alchemical FEP workflows that depend on extensive sampling for convergence. Despite this reduced computational cost, Qenergy-VM2 delivers accuracy comparable to and in some flexible systems, exceeding reported FEP+ results. This combination of speed, precision, and reliable rank-ordering enables rapid iterative design cycles compatible with realistic medicinal chemistry timelines, supporting both lead optimization and large-scale screening campaigns. Overall, Qenergy-VM2 offers a quantum-enhanced, physically grounded, and computationally efficient approach for ligand prioritization and BFEs prediction across diverse protein targets.

**Supporting Information**

All raw data files for ligands and targets are available. Residual analysis and correlation plots comparing predicted and experimental ΔG for the entire studied dataset using the Qenergy-VM2 protocol are included. Additionally, absolute error (offset-scaled) distribution histograms for each target are provided. All data are supplied in TXT format along with the article.

# Synergistic Computational Approaches for Accelerated Drug Discovery: Integrating Quantum Mechanics, Statistical Thermodynamics, and Quantum Computing

## Supplementary information

Farzad Molani[1] and Art E. Cho[1,2*]

[1] inCerebro, Co., Ltd., Seongdong-gu, Seoul, 04778, Republic of Korea

[2] Department of Bioinformatics, Korea University, 2511 Sejong-ro, Sejong, 30119, Republic of Korea

Email: artcho@korea.ac.kr

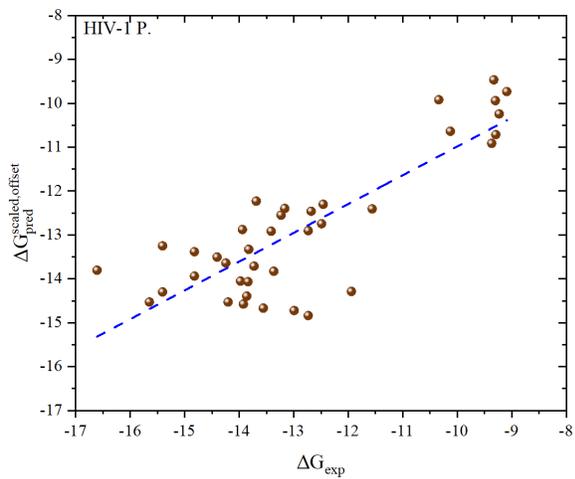
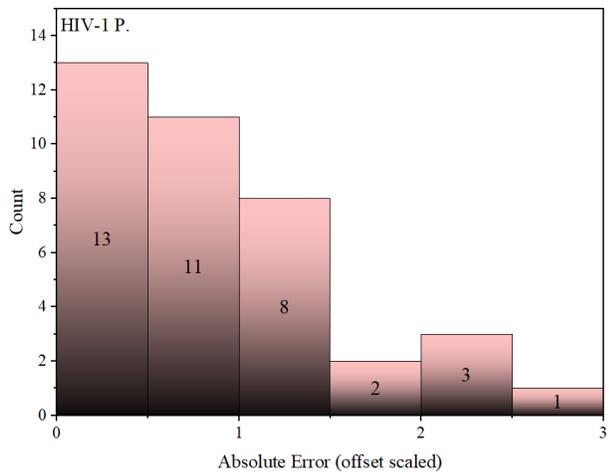
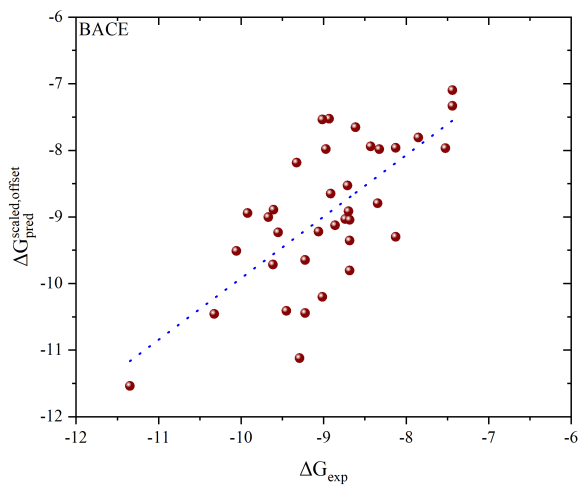
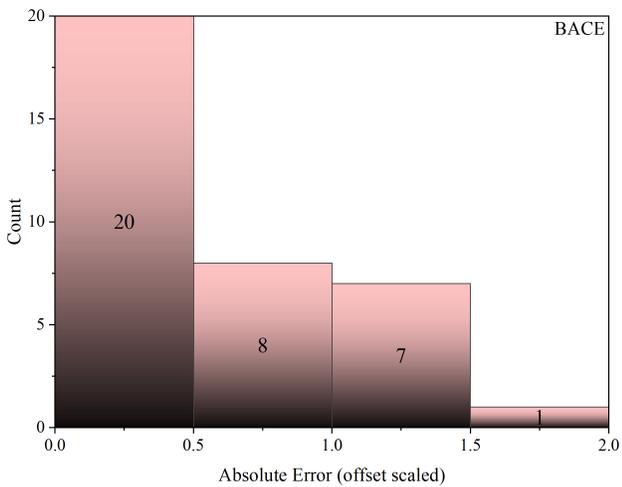
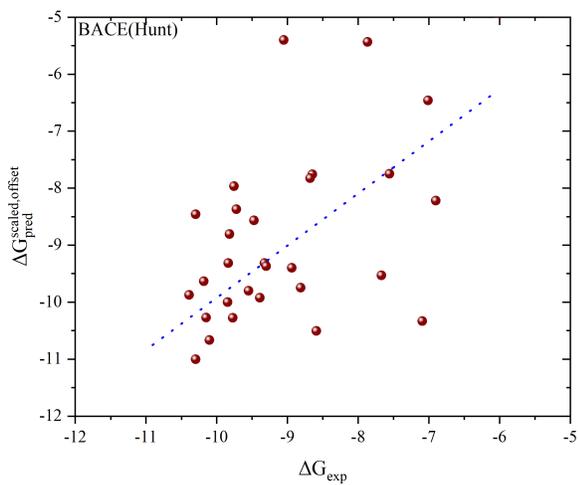
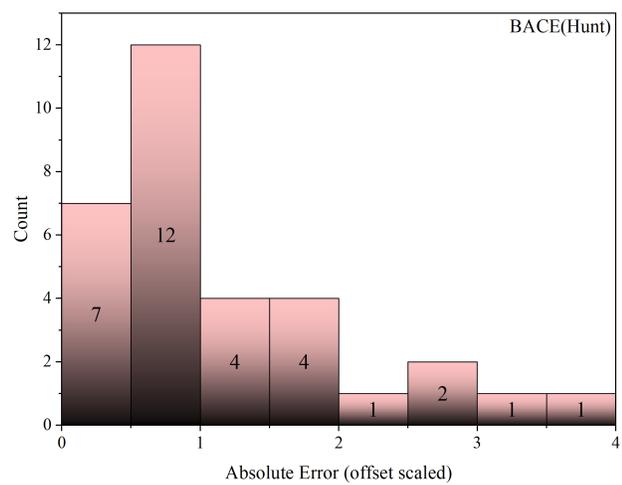

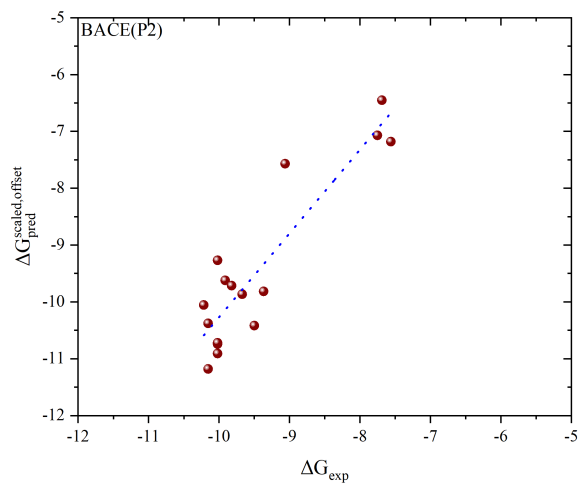
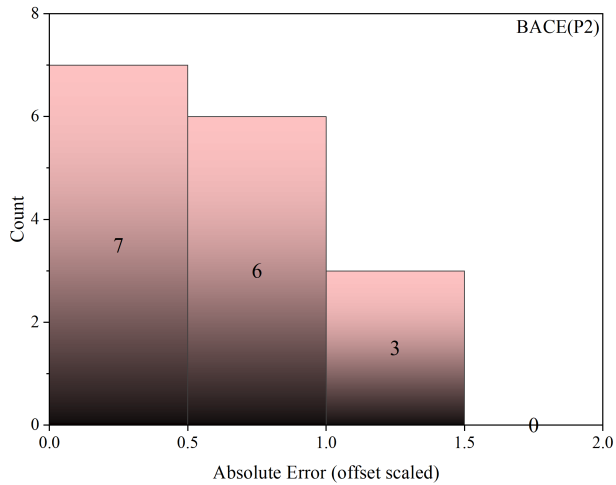
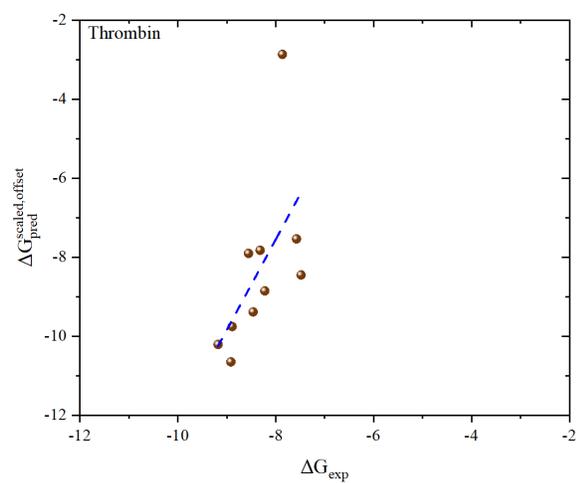
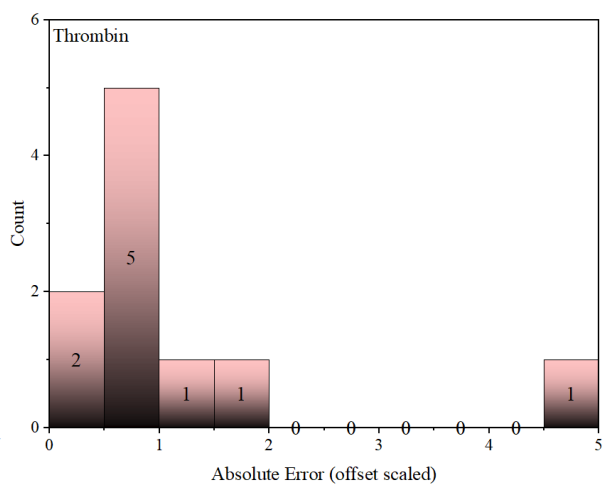
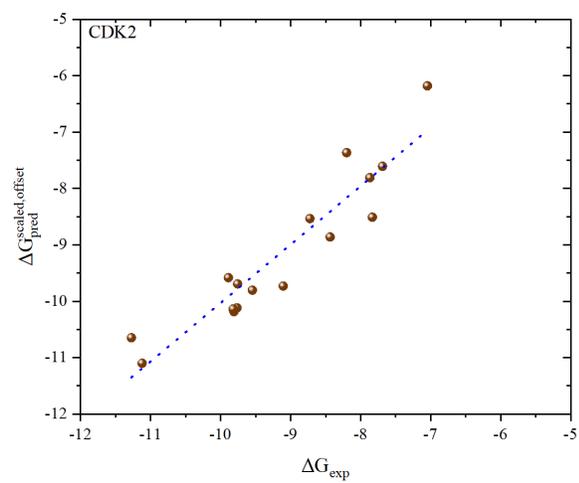
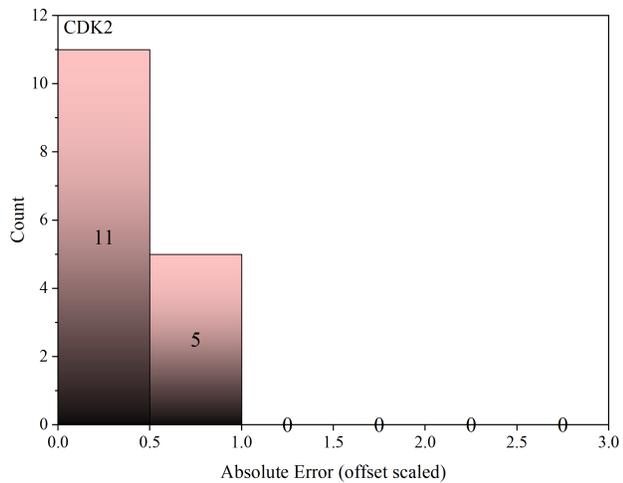

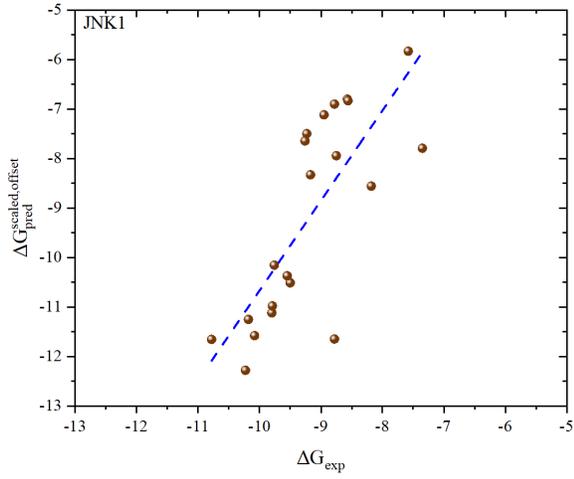
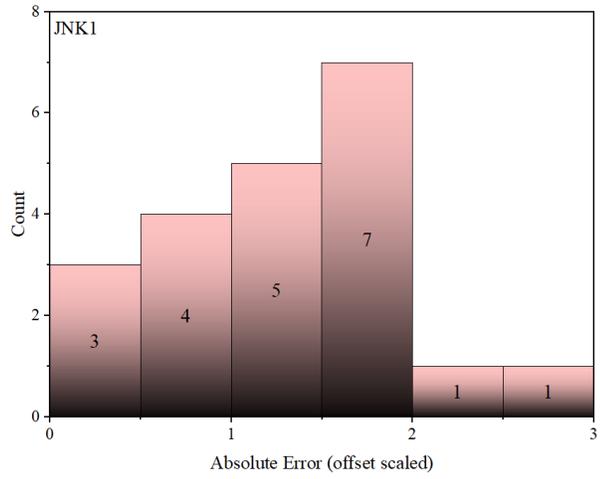
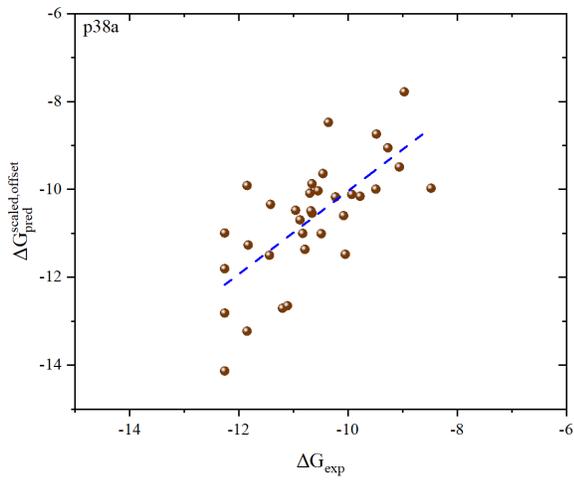
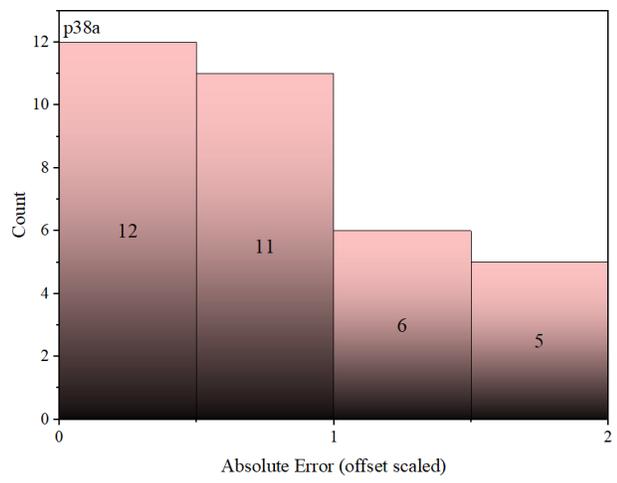
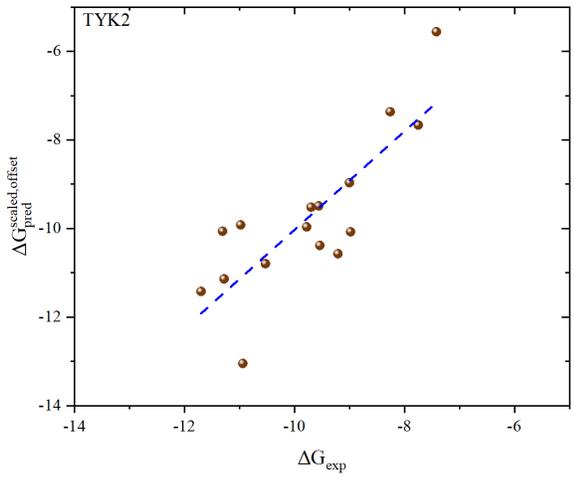
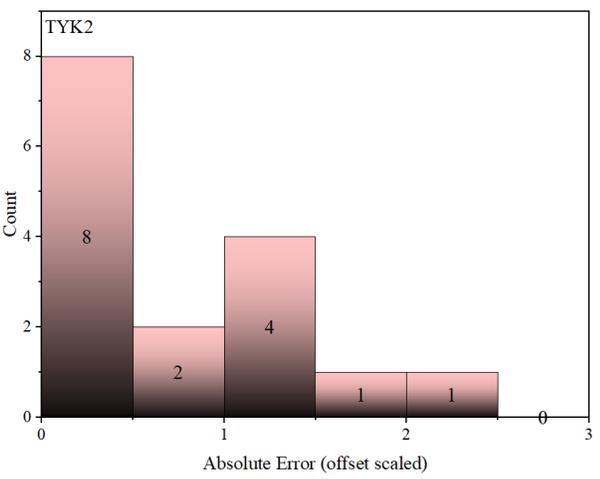

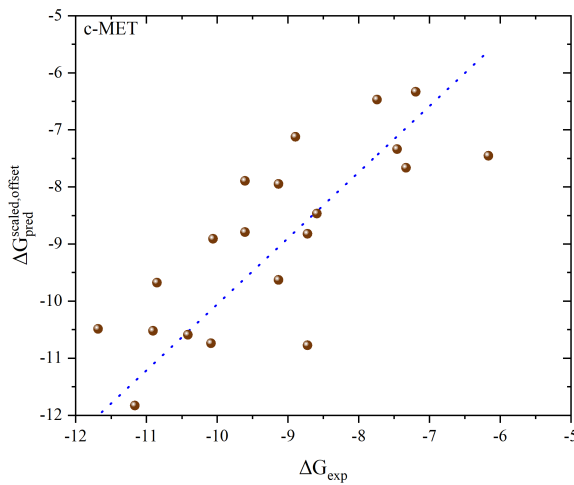
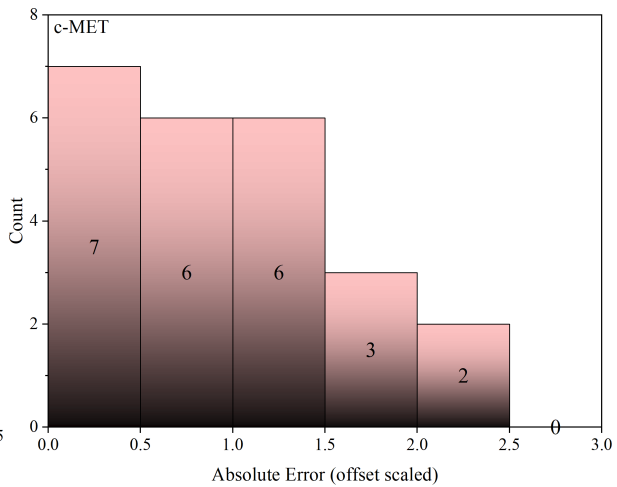
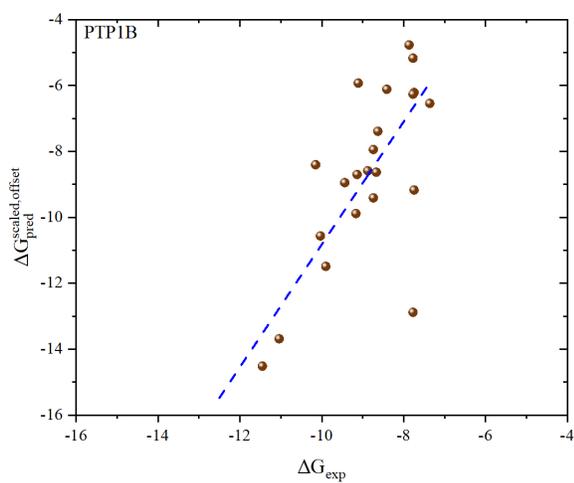
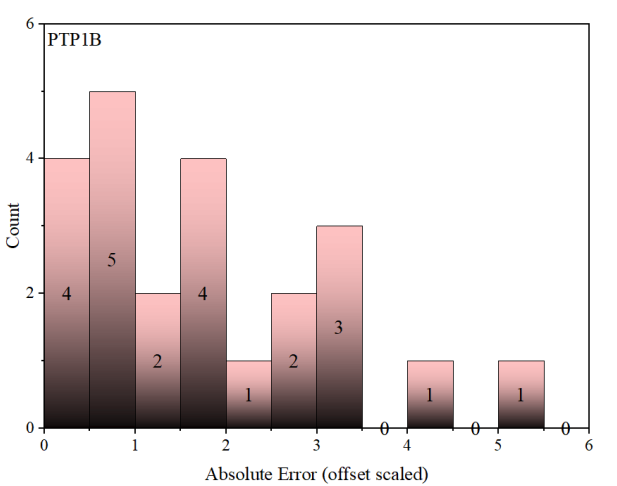
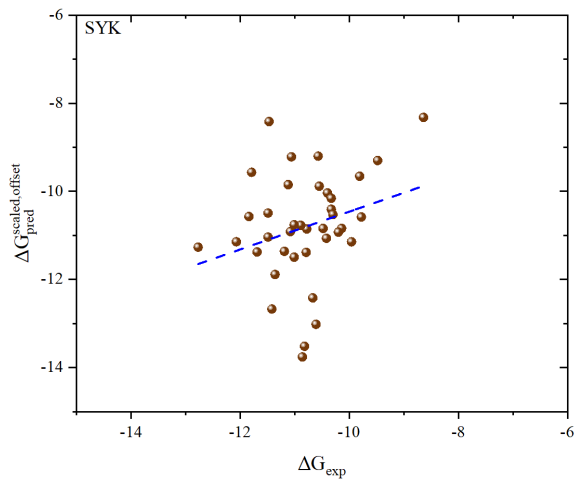
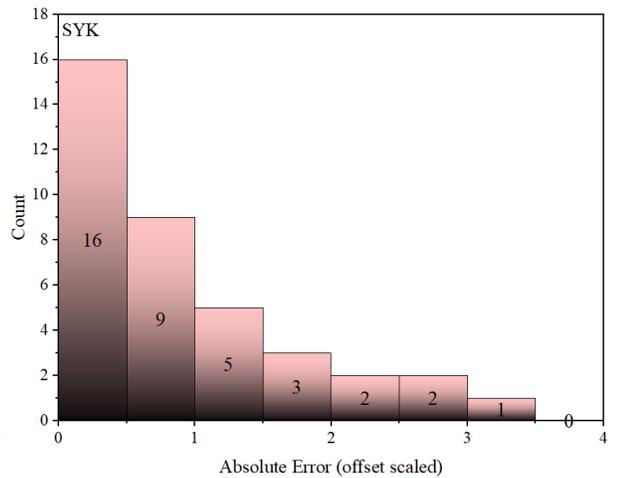

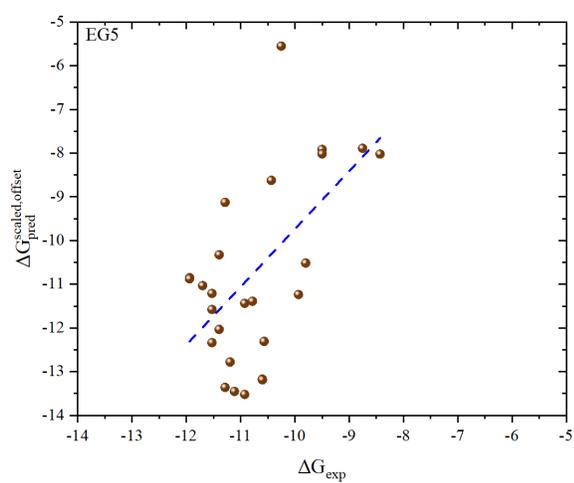
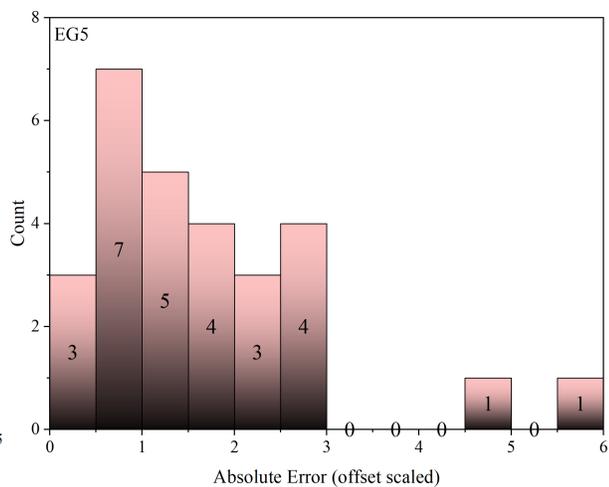
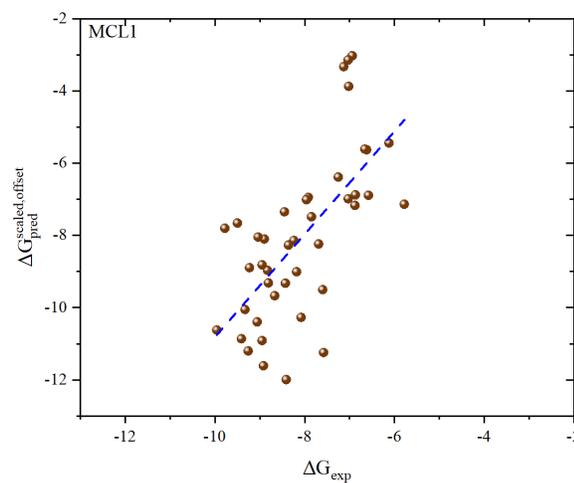
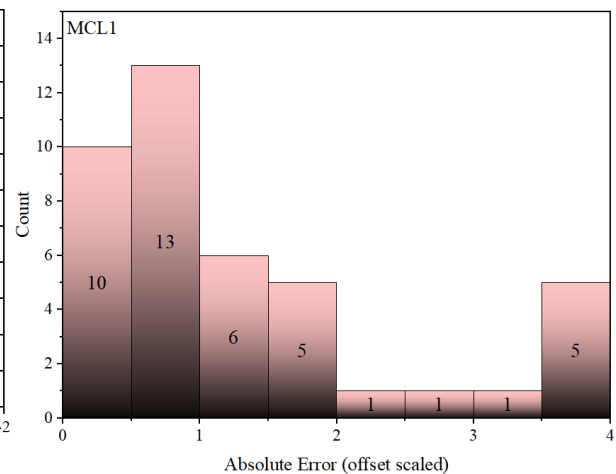
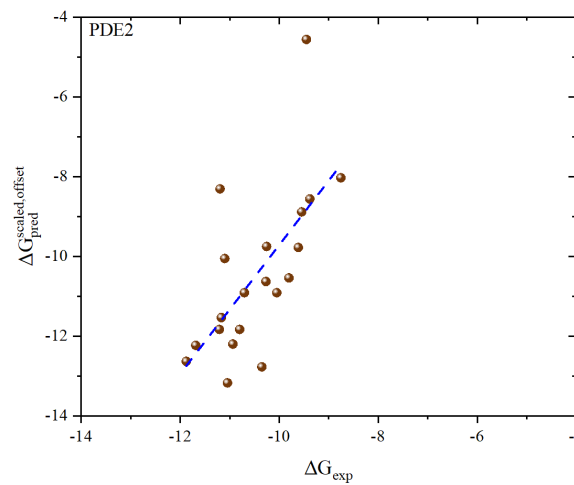
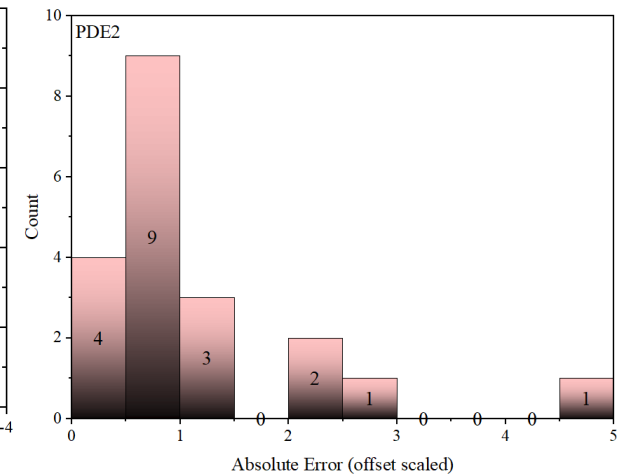

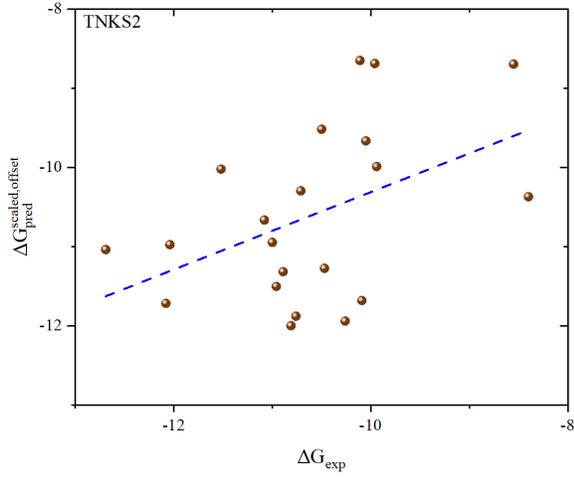
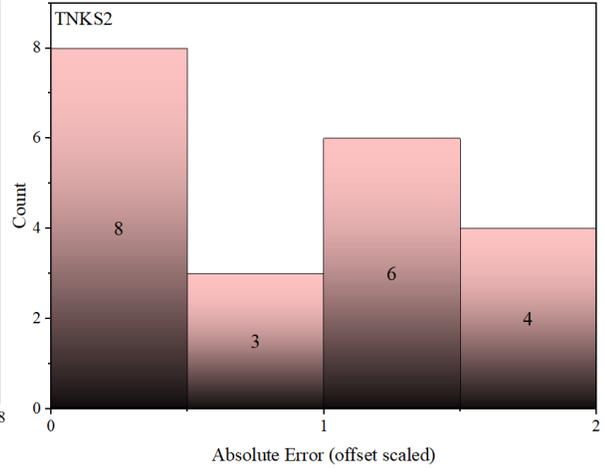
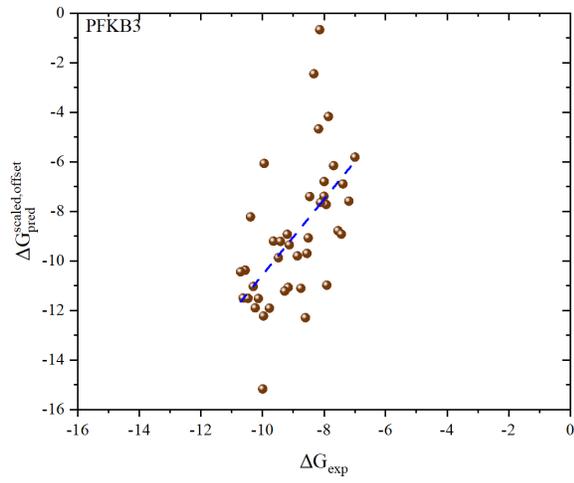
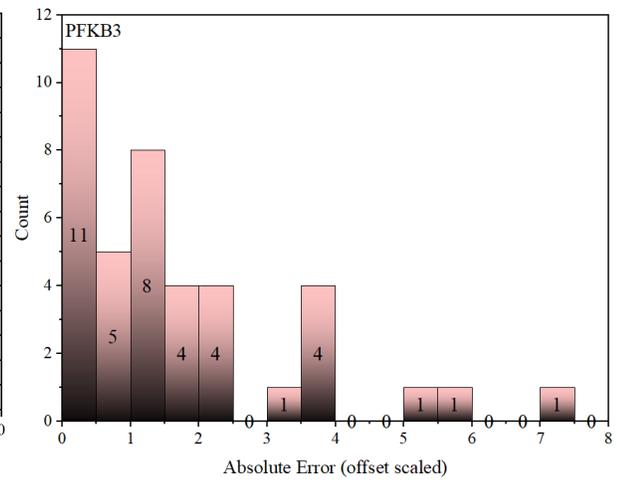
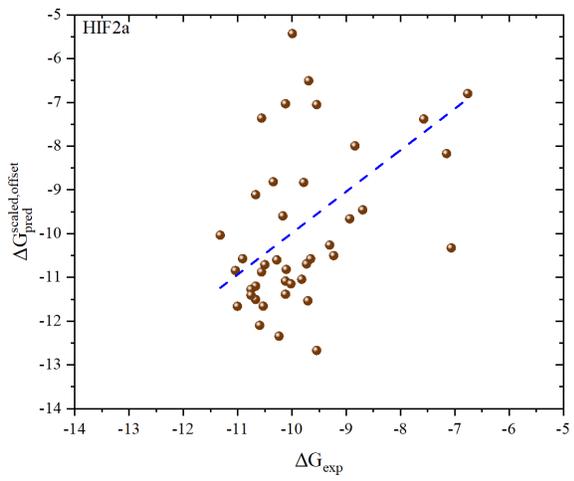
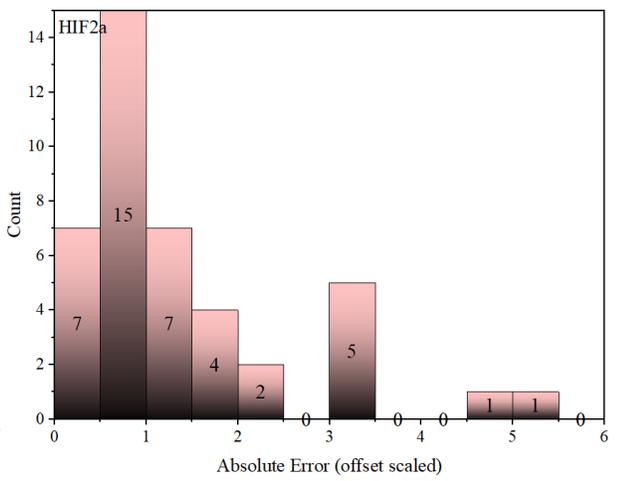

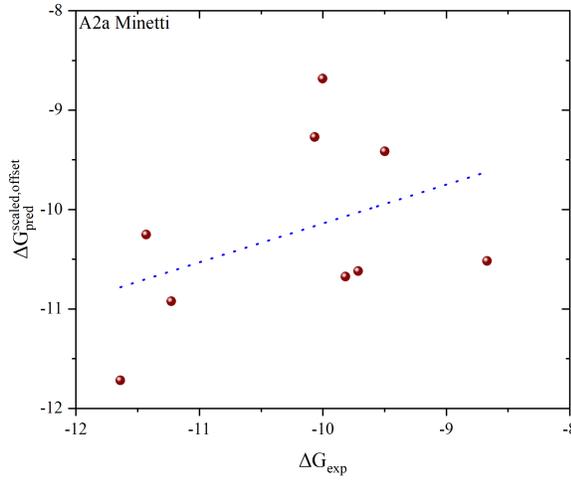
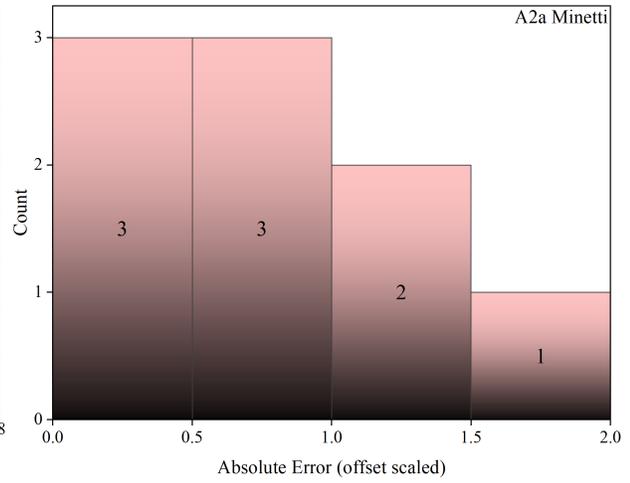
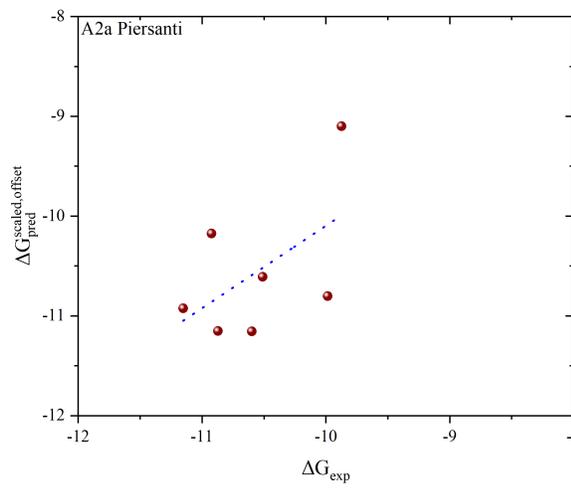
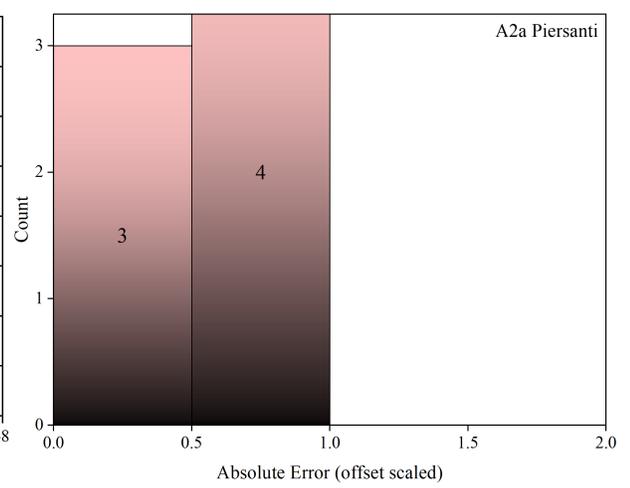
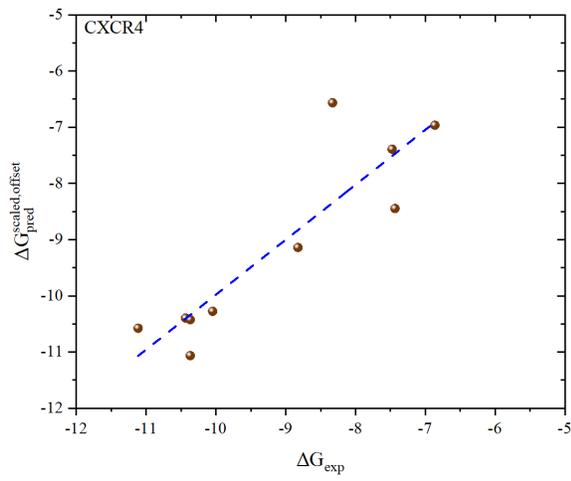
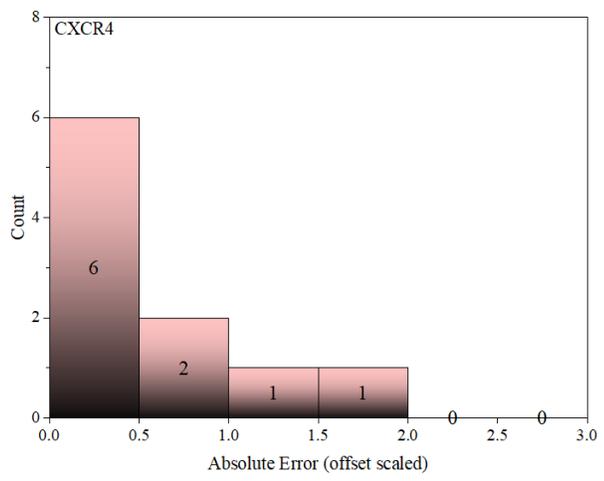

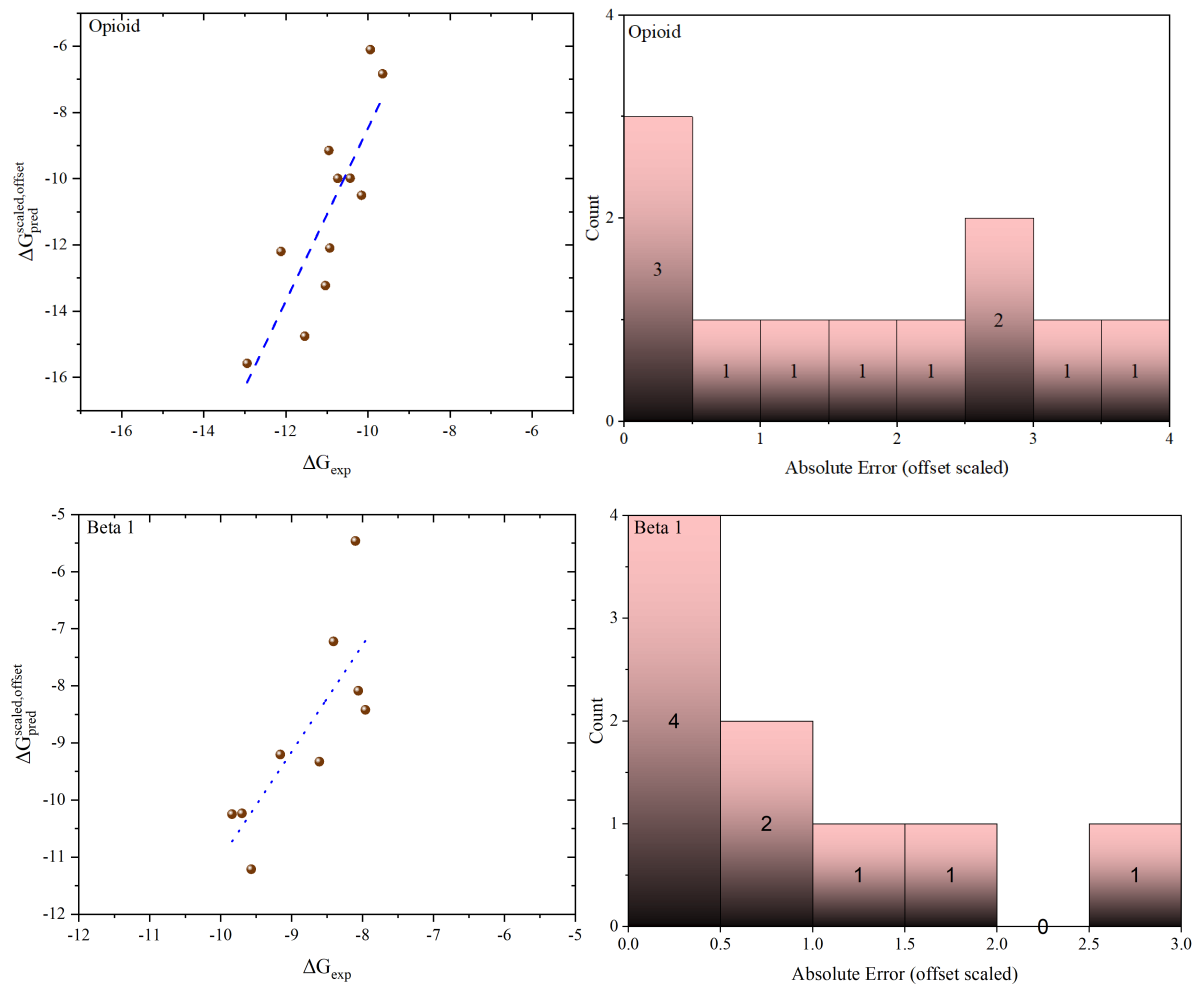

**Supplementary Figure 1** Correlation plot of predicted and experimental ΔG for whole studied dataset for Qenergy-VM2 protocol. Absolute error (offset scaled) distribution histogram for each target showed as well.

### Statistical Analysis

The consistency of response data generated by Qenergy-VM2 was evaluated using residual plot analysis with a focus on linear overcut. Residuals, which represent the differences between observed and predicted values for each data point, were analyzed. Residual plots were used to check three critical assumptions for the validity of computational runs: normality, constant variance, and independence. These assumptions were validated to assess linear overcut as the specific response factor. In residual plots, if points are randomly scattered around the horizontal axis, it indicates that a linear regression model is appropriate; otherwise, a non-linear model may be more suitable. The residual plots in Supplementary Figure 2 illustrate three distinct patterns. The first plot (Supplementary Figure 2(a)) shows residuals versus the independent variable, where a random or patternless distribution indicates a good fit for a linear model for the protocol. The histogram of residuals (Supplementary Figure 2(b)) helps determine whether the variance is normally distributed. A symmetrical, bell-shaped histogram centered around zero, as seen in Supplementary Figure 2b, supports the normality assumption for the protocol. This bell-curve is ideal for indicating normality. If the histogram shows that random errors are not normally distributed, it suggests potential violations of the model's assumptions. The residuals versus fitted values plot (Supplementary Figure 2(c)) offers further validation for these assumptions. Additionally, a normal probability plot of the residuals (Supplementary Figure 2(d)) can be used to verify normal variance distribution. An approximately linear plot suggests that the error terms are normally distributed for the protocol.

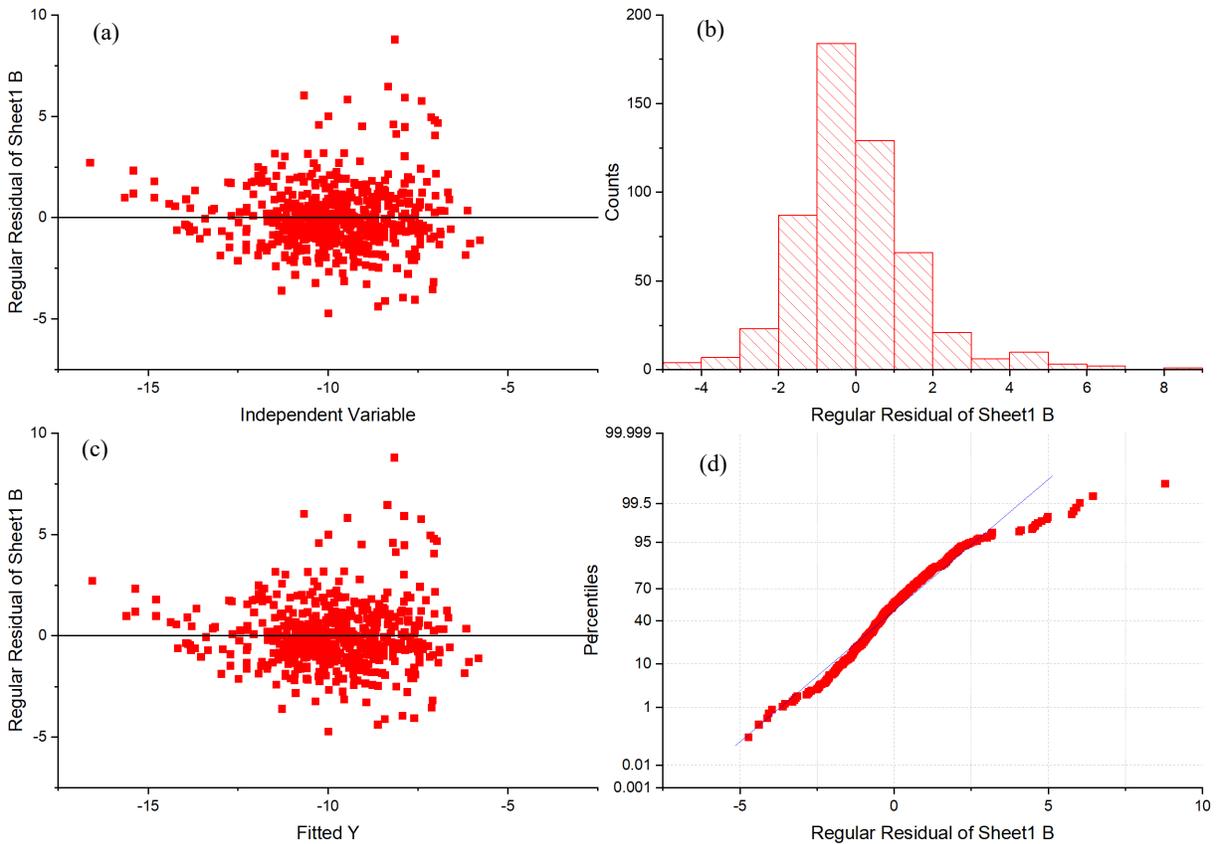

**Supplementary Figure 2** Plots showing (a) residuals versus the independent variable, (b) histogram of the residuals, (c) residuals versus the fitted values, and (d) normal probability plot of the residuals.